\documentclass{ifacconf}

\usepackage{graphicx}      
\usepackage[sort&compress,round]{natbib}        
\usepackage{notoccite}

\usepackage{epsfig} 
\usepackage{mathptmx} 
\usepackage{times} 
\usepackage{amsmath} 
\usepackage{amssymb}  
\usepackage{cancel}
\usepackage{gensymb}
\usepackage{amsfonts}
\usepackage{booktabs,tabularx}

\newcommand\numberthis{\addtocounter{equation}{1}\tag{\theequation}}
\begin{document}
\begin{frontmatter}

\title{Reset Control for Vibration Disturbance Rejection} 

\author{Erdi Aky\"{u}z, Niranjan Saikumar, S. Hassan HosseinNia}

\address{Department of Precision and Microsystems Engineering, 3ME, TU Delft\\The Netherlands}
\address{(e-mail: E.Akyuz@student.tudelft.nl, N.Saikumar@tudelft.nl, S.H.HosseinNiaKani@tudelft.nl)}

\begin{abstract}
The high tech industry which requires fast stable motion with nanometer precision continues to mainly use PID which is limited by fundamental linear control limitations. Floor vibrations as disturbance significantly affect performance and their rejection is particularly affected by these limitations. Reset control has provided a promising alternative to surpass these, while simultaneously allowing the use of industry standard loop-shaping during design. However, the reset action introduced higher order harmonics can induce unwanted dynamics and negatively influence performance. This paper investigates two reset control strategies namely (1) band-pass phase lag reduction and (2) phase compensation to reduce the negative effects of higher order harmonics. The strategies are tested on a precision positioning stage for vibration rejection and the results show that phase compensation provides better performance compared to other tested strategies.
\end{abstract}


\end{frontmatter}

\section{Introduction}
Tracking, steady-state precision and bandwidth are the primary design specifications that require constant improvement for continued excellence in the high tech industry. Meeting sufficient stability margins and robustness is assumed to be part of all controller designs. These requirements are especially eminent in photolithography machines, AFMs, white light interferometers and other precision instruments. However, these systems do not operate in isolation and are affected by external disturbances. Disturbance rejection is another specification which has to be met to reach desired performance in the real world. Floor vibrations, an external disturbance, is ubiquitous in practice and vastly hinders the system performance. Although their intermittent character makes it impossible to model them in advance, they are present mainly at frequencies ($0.5$ - $30$~Hz) which are below the bandwidth from the perspective of high-speed motion systems under consideration [\cite{ryan2017optical}]. Effective vibration disturbance rejection through good feedback controller design is an absolute necessity.

PID which is widely used in the precision industry is unable to meet the increasing demands of bandwidths and precision [\cite{futurePID}]. Double integrators are generally used as part of the controller to achieve nanometer precision and good disturbance rejection [\cite{mao2003double}]. Error minimization is achieved at low frequencies from increased open-loop gain. However, this also decreases the phase margin and reduces the overall robustness of the system; resulting in higher overshoot and increased settling times. This conflict between disturbance rejection \& tracking performance on one side with stability margins and robustness on the other is a direct result of linear control limitations. Waterbed effect explains that improved disturbance rejection in a range of frequencies results in deterioration of the same in other regions [\cite{MSDbook}]. From Bode's gain-phase relationship, increasing gain at low frequencies has a negative influence on phase, compromising robustness. Simultaneous match of required disturbance rejection and stability specifications requires that these limitations of linear controllers are overcome, with the introduction of nonlinear behaviour like reset.

JC Clegg proposed reset in 1958 with the reset integrator to overcome the aforementioned limitations [\cite{Clegg1958}]. The describing function evaluation of this so-called Clegg integrator (CI) shows 51.9$\degree$ less phase lag and gain characteristics 1.62 times higher compared to linear integrator and could hence provide reduced overshoot and improved stability. This property has been utilized in literature to increase the bandwidth which improves disturbance rejection in general. However, very few works specifically for improved disturbance rejection using reset exist in literature to the best of authors' knowledge. \cite{Guo2009} achieve good midfrequency disturbance rejection through reset for a narrowband. However, broadband rejection is not explored. Further, reset action also creates higher order harmonics and induces unwanted dynamics resulting in performance deterioration which cannot be analysed using describing function. Some methods exist in literature as both means to tune reset elements and reduce these harmonics. Partial reset, PI+CI [\cite{Banos2012PICL}], reset band [\cite{Vidal2008}], time regularization [\cite{zheng2007improved}] provide means to counter this issue, but rather as a trade-off between linear and reset control. Further, tuning for these methods is not clear and they are ineffective regarding improvement to disturbance rejection.

This paper studies and presents techniques for the use of reset for improved vibration disturbance rejection and trajectory tracking resulting in improved performance. Preliminaries are provided in Section \ref{Prelim} with the two different approaches for reducing higher order harmonics and improving performance presented in Section \ref{strategies}. The results and analysis are presented in Section \ref{validation} followed by the conclusions in Section \ref{Conclude}.

\section{Preliminaries}
\label{Prelim}
\subsection{Vibration rejection in the frequency domain}
Floor vibrations exist in the range of 0.5 to 30 Hz, i.e., at frequencies lower than bandwidth for high-speed motion systems. In systems such as AFMs, the vibrations act as a direct disturbance force on the mass being controlled and creates errors, requiring good disturbance rejection. However, in the case of in-line metrology platform of \cite{saathof2017integrated}, vibrations act on the workpiece (placed on the platform) creating relative motion from perspective of inspection tool. Since the tool is being controlled, this results in an unknown reference which has to be tracked and requires the controller to have good reference tracking characteristics. Since vibrations are not known in advance, well-designed feedforward controllers which are used in precision motion systems cannot be used in this case. 

\begin{figure}
	\centering
	\includegraphics[trim = {1cm 0.5cm 0cm 0.5cm},width=\linewidth]{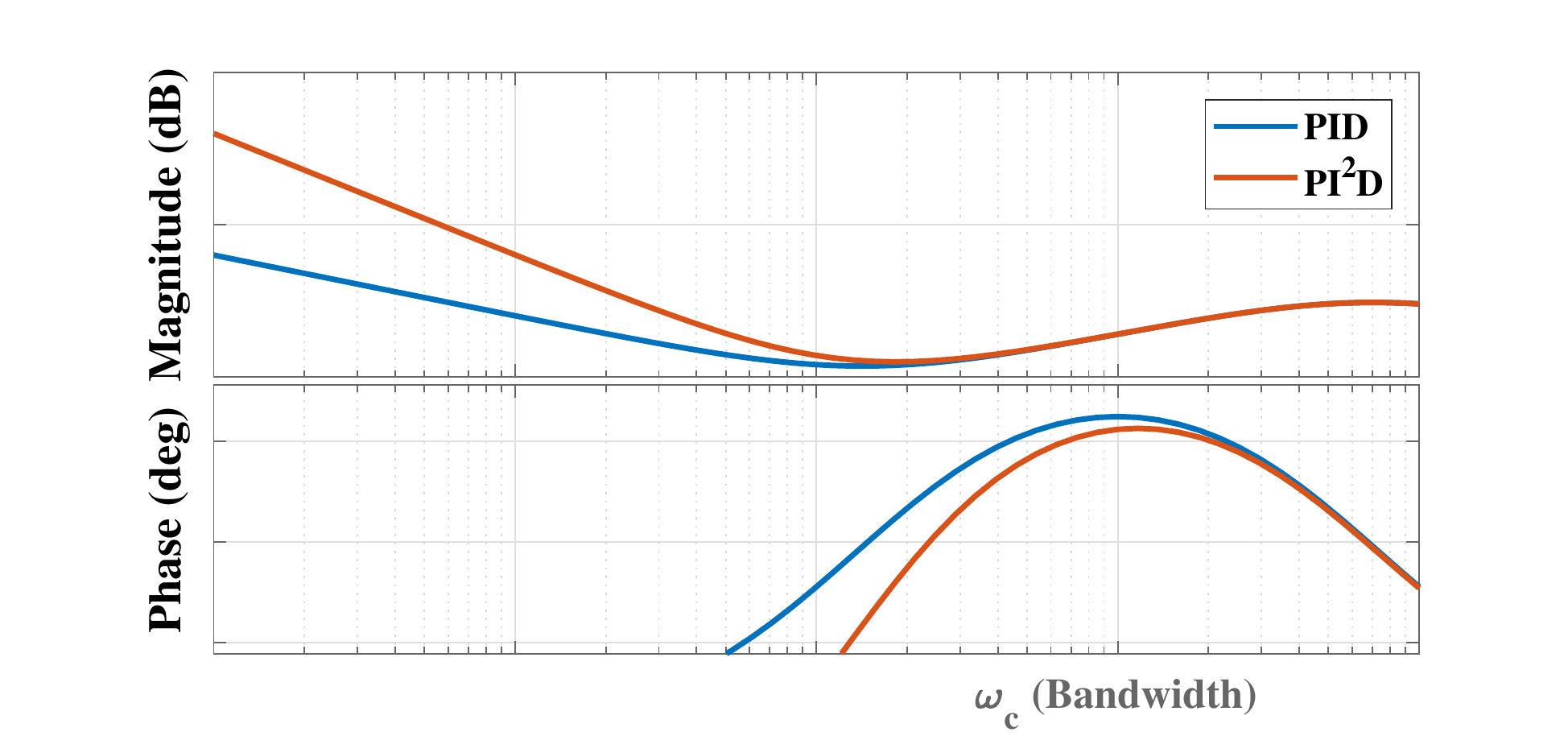}
	\caption{Frequency domain comparison of using single and double integrators for improved disturbance rejection and reference tracking.}
	\label{fig:freqdomain}
\end{figure}%

Both reference tracking and disturbance rejection are improved through the use of integrator action increasing the open-loop gain as shown in Figure \ref{fig:freqdomain}. While anti-notch filters can be used to increase gain in narrowband, integrators provide the freedom for broadband disturbance rejection. The use of double integrator further improves rejection as seen in the same figure but at the cost of reduced phase margin. Bode's gain-phase relationship which explains this trade-off can be overcome through the use of nonlinearity in the form of reset.

\subsection{Reset control definition}
A reset controller is a type of impulsive system with a linear base controller and a reset mechanism. This mechanism provokes the control signal to change when the resetting condition is satisfied [\cite{Banos2012}]. This action overcomes the linear control limitations and provides the phase advantage noted by Clegg. 

A reset controller (consisting of resetting part and non-resetting parts) combined with the plant in series (resulting in open-loop) can be represented as
\begin{align*}
\label{eq:reset}
\dot{x}&=Ax+Be,&x,e,t\notin\mathcal{M}\\
x(t^+)&=A_R x,&x,e,t\in\mathcal{M} \numberthis \\
y&=Cx+De& 
\end{align*}

where $e$ is input error signal, $y$ is plant output, $x:[x_R^T,x_p^T]^T$ is the state vector with $x_R$ and $x_p$ denoting reset controller and plant states respectively. $x_R(t)=[x_r^T\;x_{nr}^T]^T$ where $x_r \in \mathbb{R}^{n_r\times 1}$ and $x_{nr} \in \mathbb{R}^{n_{nr}\times 1}$ are states of the reset element ($C_r$) and non-resetting linear controller ($C_{nr}$) respectively. The first and third equations show the continuous flow dynamics with the second equation showing the jump mode. This jumping action introduces nonlinearity to the controller. The dynamics of jump are governed by resetting matrix $A_R$. $A_R=[{A_{\rho}}_{n_r\times n_r} 0;0\; I_{n_{nr}}]$. While various reset conditions have been studied in literature, the most popular condition due to advantages in analysis and tuning used in literature is zero crossing of error input, i.e., $e(t) = 0$ and this is used in this paper.

\subsection{Describing Function}
Due to the nonlinear nature of reset controllers, describing function analysis (DF) is used to study the frequency domain behaviour. DF is based on a quasi-linearization and sinusoidal inputs are generally considered for this purpose [\cite{DFbook}]. Since ground vibrations are sinusoidal in nature, this approach holds water. The analytical equation to determine DF for reset systems with $e(t) = 0$ as reset condition is provided by \cite{Guo2009} as
\begin{align}
\label{eq:DF}
G(j\omega)=C^T(j\omega I-A)^{-1}(I+j\Theta_D(\omega))B+D
\end{align}
where
\begin{align}
\Theta_D(\omega)&\buildrel{\Delta}\over{=}-\frac{2\omega^2}{\pi}\Delta(\omega)(\Gamma_R(\omega)-\Lambda^{-1}(\omega))
\end{align}
\begin{align*}
\Lambda (\omega) &\buildrel{\Delta}\over{=}\omega^2+A^2\\
\Delta(\omega)&\buildrel{\Delta}\over{=}I+e^{\frac{\pi}{\omega}A}\\
\Delta_R(\omega)&\buildrel{\Delta}\over{=}I+A_Re^{\frac{\pi}{\omega}A}\\
\Gamma_R(\omega)&=\Delta_R^{-1}(\omega)A_R\Delta(\omega)\Lambda^{-1}(\omega)
\end{align*} 

\begin{figure}
	\centering
	\includegraphics[width=\linewidth]{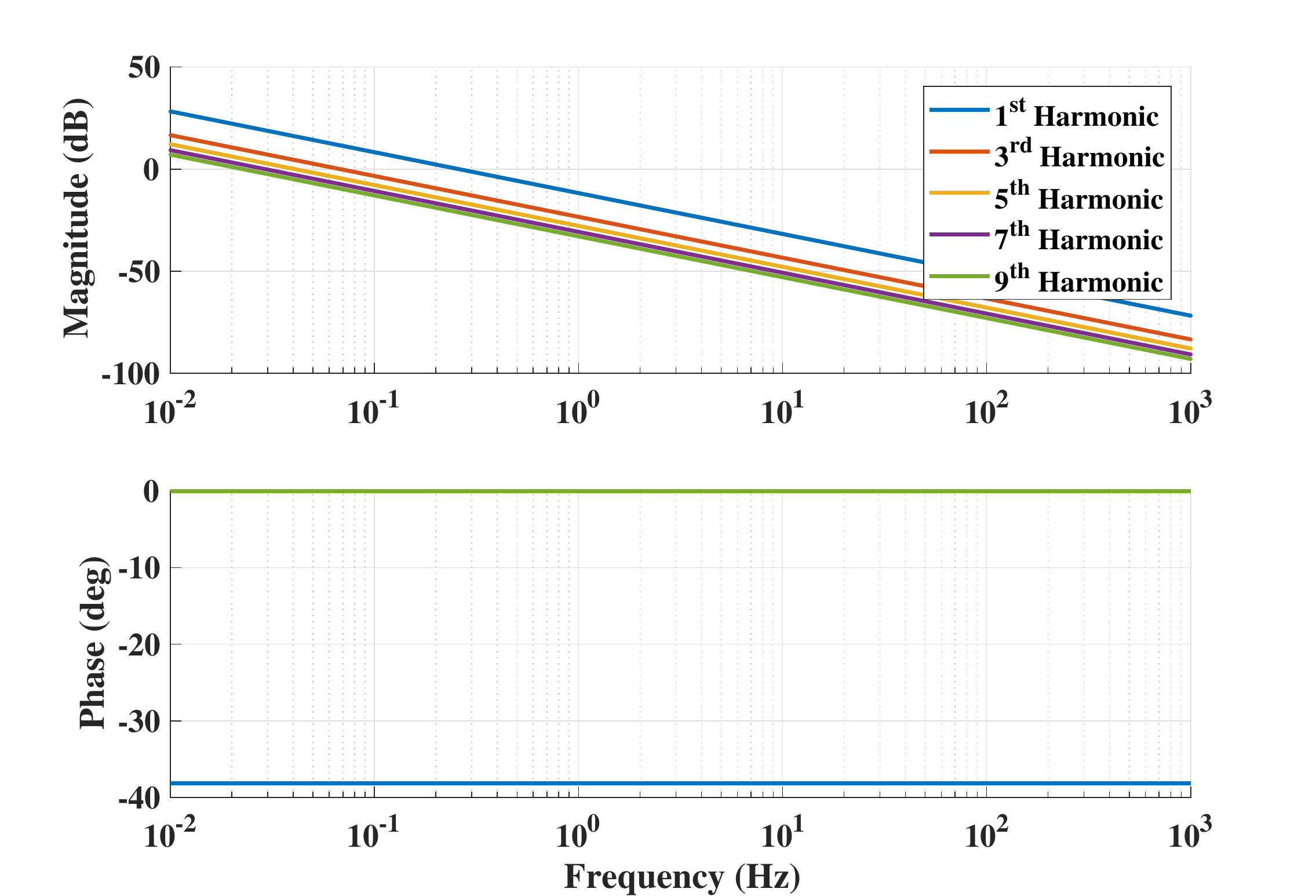}
	\caption{Frequency behaviour of Clegg integrator showing higher order harmonics up-to 9\textsuperscript{th} order. Note that the even harmonics are zero and hence not shown.}
	\label{fig:clegghigh}
\end{figure}

Describing function analysis considers only the first harmonic and neglects the higher order harmonics with the assumption that these can be neglected. However, the idea of higher-order sinusoidal input describing functions (HOSIDFs) for nonlinear elements introduced by \cite{NUIJ20061883} has been extended to reset controllers by \cite{Kars} with the analytical equations for HOSIDFs provided as
\begin{align*}
G(j\omega,n)=&\begin{cases}
\frac{-2\omega^2C}{j\pi}(A-j\omega nI)^{-1}\Delta(\omega)\big[\Gamma_R(\omega)-\Lambda^{-1}(\omega)\big]B, \numberthis \label{eq:HOSIDF}\\
\;\;\;\;\;\;\;\;\;\;\;\;\;\;\;\;\;\;\;\;\;\;\;\;\;\;\;\;\;\;\;\;\;\;\;\;\;\;\;\;\;\text{for odd }n\geq 2\\
0,\;\;\;\;\;\;\;\;\;\;\;\;\;\;\;\;\;\;\;\;\;\;\;\;\;\;\;\;\;\;\;\;\;\;\;\;\;\;\text{for even }n\geq 2
\end{cases}
\end{align*}
where $n$ denotes the order of the harmonics. Eqns. (\ref{eq:DF}) and (\ref{eq:HOSIDF}) are used to plot the frequency behaviour of a Clegg integrator (CI) showing harmonics up-to the 9\textsuperscript{th} order in Fig. \ref{fig:clegghigh}. The combination of these tools provides us with a better picture of the frequency domain behaviour of reset controllers in open-loop. From Fig. \ref{fig:clegghigh}, while it is seen that the magnitude of higher harmonics is lesser than that of first for CI, they cannot be completely neglected as is done in literature.

The stability properties of reset systems and the conditions for stability have been extensively studied in literature. We refer the readers to the work of \cite{beker2004fundamental} for $H_\beta$ conditions and also to the work of \cite{nevsic2008stability} using piece-wise Lyapunov equations.

\section{Reset strategies for improved vibration rejection}
\label{strategies}
In this section, two different approaches are considered for the use of reset to improve vibration rejection. These are 1) band-pass phase lag reduction and 2) phase compensation.

\subsection{Band-pass phase lag reduction}
\begin{figure}
	\centering
	\includegraphics[width=.9\linewidth]{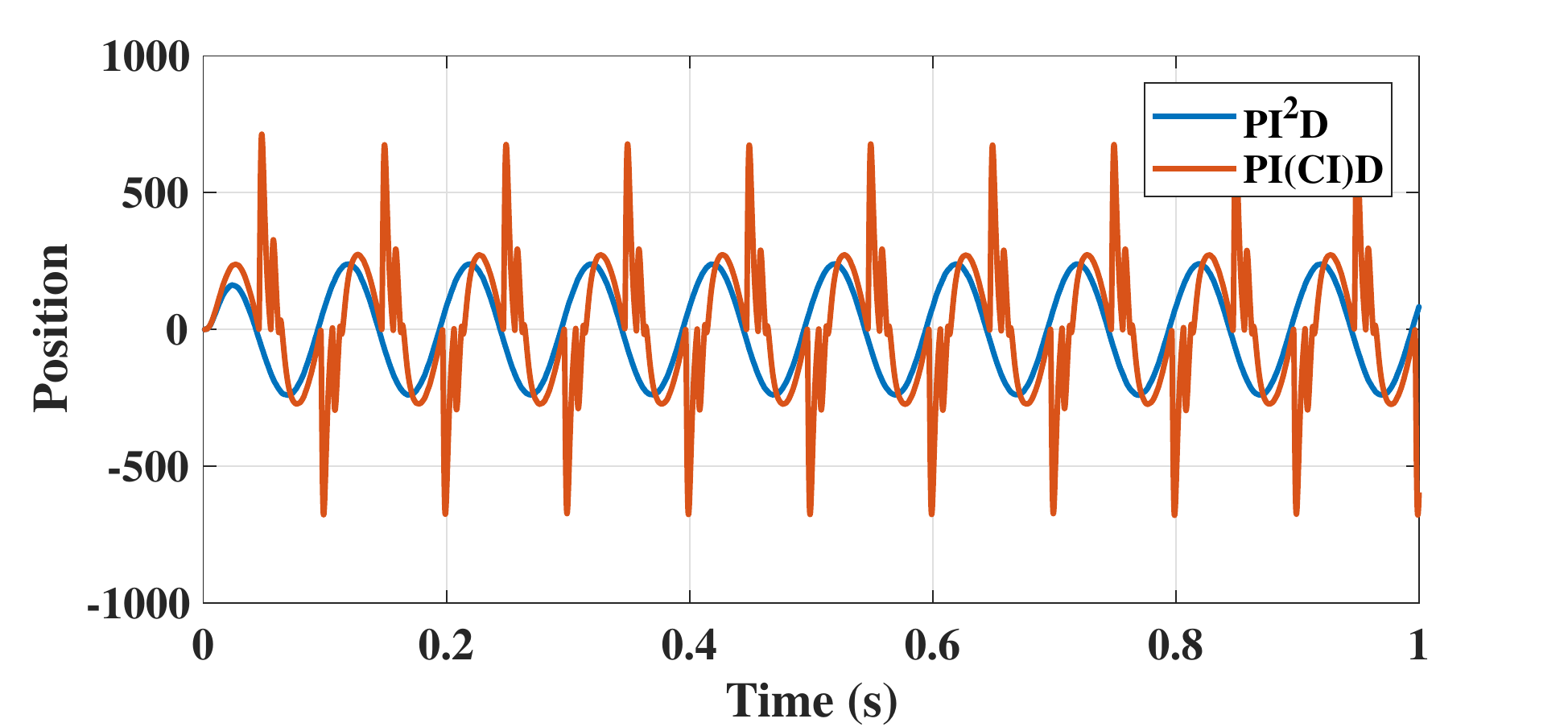}
	\caption{Response to 10 Hz sinusoidal input disturbance.}
	\label{allPIiDdisturbance}
\end{figure}

The conventional use of reset in literature has been for phase lag reduction. Starting from CI, reset has been introduced to lag filters since they have lesser phase lag compared to their linear counterparts. In our case of PI\textsuperscript{2}D, one of the integrators is replaced by CI resulting in what is referred to as PI(CI)D. The use of PI(CI)D results in a phase margin increase according to DF analysis. A simulation study on a mass system is conducted with PI\textsuperscript{2}D and with one of the integrators replaced by CI to get PI(CI)D. The response of this closed-loop system to a 10 Hz sinusoidal disturbance shown in Fig. \ref{allPIiDdisturbance} clearly shows that the higher order harmonics which are not analysed with DF negatively affect system performance.

\begin{figure}
	\centering
	\includegraphics[width=.6\linewidth]{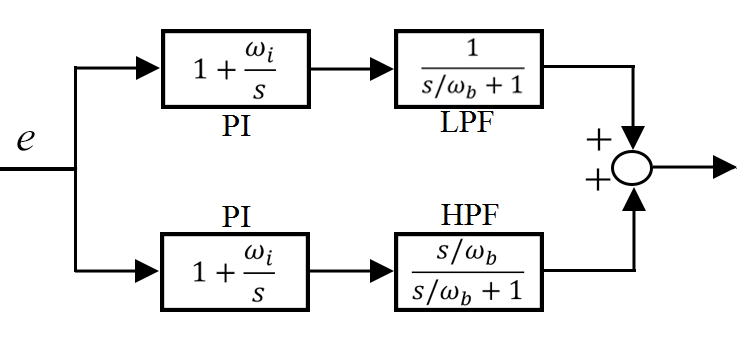}
	\caption{Band-Pass resetting part which is to replace CI in PI(CI)D}
	\label{fig:sub3}
\end{figure}

As noted earlier, a HOSIDF analysis of CI in Fig. \ref{fig:clegghigh} shows that the higher order harmonics while smaller in magnitude than the first harmonic, are not completely negligible and clearly have an effect on the closed-loop system performance. We hypothesize that the attenuation of these harmonics in the required frequency range can result in performance improvement. Since this work looks at rejecting vibrations which are in the frequency range of 0.5 to 30 Hz, we propose a band-pass reset approach to verify this hypothesis. The idea is to use the reset action only in the required frequency range while reverting to a linear configuration in other ranges. This is achieved by utilizing two parallel branches with one housing a low pass filter and another a high pass filter as shown in Fig. \ref{fig:sub3}. The figure shows the potential replacement for the CI part of PI(CI)D, while the rest of the controller (linear PID) is assumed to be in series such that the output of Fig. \ref{fig:sub3} is fed as input to PID.

Since the cut-off frequencies of both LPF and HPF filters are the same as seen in Fig. \ref{fig:sub3}, the result of the configuration shown is a linear PI filter $(1 + \omega_i/s)$. This configuration provides us with different possibilities to introduce reset into the controller where we can obtain the advantage of reset while at the same time also have limited control over the magnitude of the higher order harmonics. As this is a preliminary study, the 4 filters of the band-pass configuration are considered separately for the introduction of reset. For sake of brevity, abbreviations are used for the 4 possibilities considered as given below. Note that in all cases, only one the filters is reset while the others are designed to have linear behaviour. Also in all cases, this band-pass reset element is in series with a linear PID controller.

\begin{itemize}
	\item C(IbLPF): Resetting Integrator in series with LPF 
	\item C(IbHPF): Resetting Integrator in series with HPF 
	\item C(HPF): Resetting HPF 
	\item C(LPF): Resetting LPF 
\end{itemize}

It is self-evident that in the case of IbLPF, that the reset action is used at low frequencies below the cut-off of the low pass filter, while the opposite is true in the case of IbHPF. Further, resetting of LPF and HPF results in characteristics as yet unexplored in literature for this purpose. The effect that these configurations have on the presence of higher order harmonics and on closed-loop system performance is explained in detail in Section \ref{validation}.

\subsection{Phase compensation}
Reset has mainly been used in literature for its phase lag reduction. However works in \cite{arun2018, saikumar2018constant, saikumar2019resetting} show that broadband phase compensation could be achieved with reset where phase lead is introduced indirectly, through the smart design of the Constant in gain Lead in phase (CgLp) element. With CgLp, a reset lag filter is used in series with a linear lead filter with both filters having the same cut-off frequency. In the linear case, the lag filter would cancel out the lead filter resulting in unity gain with a phase of zero. However, since the lag filter is reset in CgLp, it has less phase lag while having similar gain characteristics as the linear one. This results in unity gain while introducing phase lead into the system.

The design of CgLp element is provided as combination of the below given two elements in series:
\begin{align}
R=\frac{1}{\cancelto{}{s/\omega_{r\alpha} +1}}&& \text{and}&&L=\frac{s/\omega_r+1}{s/\omega_f+1}
\end{align}
where R and L denote first order reset lag filter and first order linear lead filter respectively.
This CgLp filter provides phase compensation mainly in the range [$\omega_r$,$\omega_f$], where $\omega_f>\omega_{r\alpha},\omega_r$. The cut-off frequency of R is not exactly the same as L, but instead also takes in a correction factor to accommodate for the small change in gain behaviour seen with the introduction of reset. This is explained in greater detail in \cite{saikumar2018constant}.

This phase lead which can be achieved through CgLp can be used to compensate for the phase margin difference seen between PID and PI\textsuperscript{2}D, where CgLp is introduced in series with PI\textsuperscript{2}D to form CgLp-PI\textsuperscript{2}D. Further, we should note again that CgLp is designed to provide phase lead without altering the gain characteristics of the open-loop. However, this analysis of CgLp is also based on DF analysis and higher order harmonics which can potentially deteriorate performance are introduced even by CgLp. The main advantage of using CgLp is it can be designed to provide phase lead in the desired frequency range which is in the region of bandwidth. This ensures that the higher order harmonics introduced are also limited to this smaller frequency range and hopefully have reduced negative effects.

In terms of controller configurations, we have the traditional designs of PID, PI\textsuperscript{2}D, PI(CI)D. Under phase lag reduction techniques where we introduce the band-pass reset approach, we have C(IbLPF), C(IbHPF), C(LPF) and C(HPF) as four potential candidates and with phase compensation we have CgLp-PI\textsuperscript{2}D. All these 8 controllers are analysed in terms of DF and HOSIDF and also implemented on a precision stage and tested for various vibration rejection performance metrics in the next section.

\section{Experimental validation}
\label{validation}
\begin{figure}
	\centering
	\includegraphics[width=0.65\linewidth]{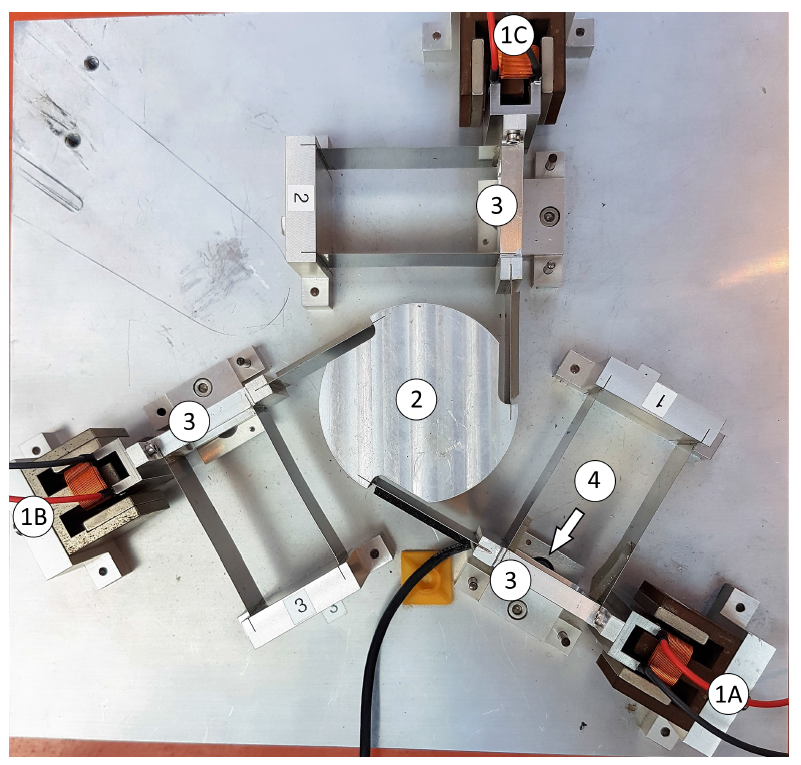}
	\caption{Precision planar positioning `Spyder' stage. Voice coil actuators 1A, 1B and 1C control 3 masses (indicated as 3) which are constrained by leaf flexures. The 3 masses are connected to central mass (indicated by 2) through leaf flexures. Linear encoders (indicated by 4) placed under masses '3' provide position feedback.}
	\label{fig:setup}	
\end{figure}

\begin{figure}
	\centering
	\includegraphics[width=0.9\linewidth]{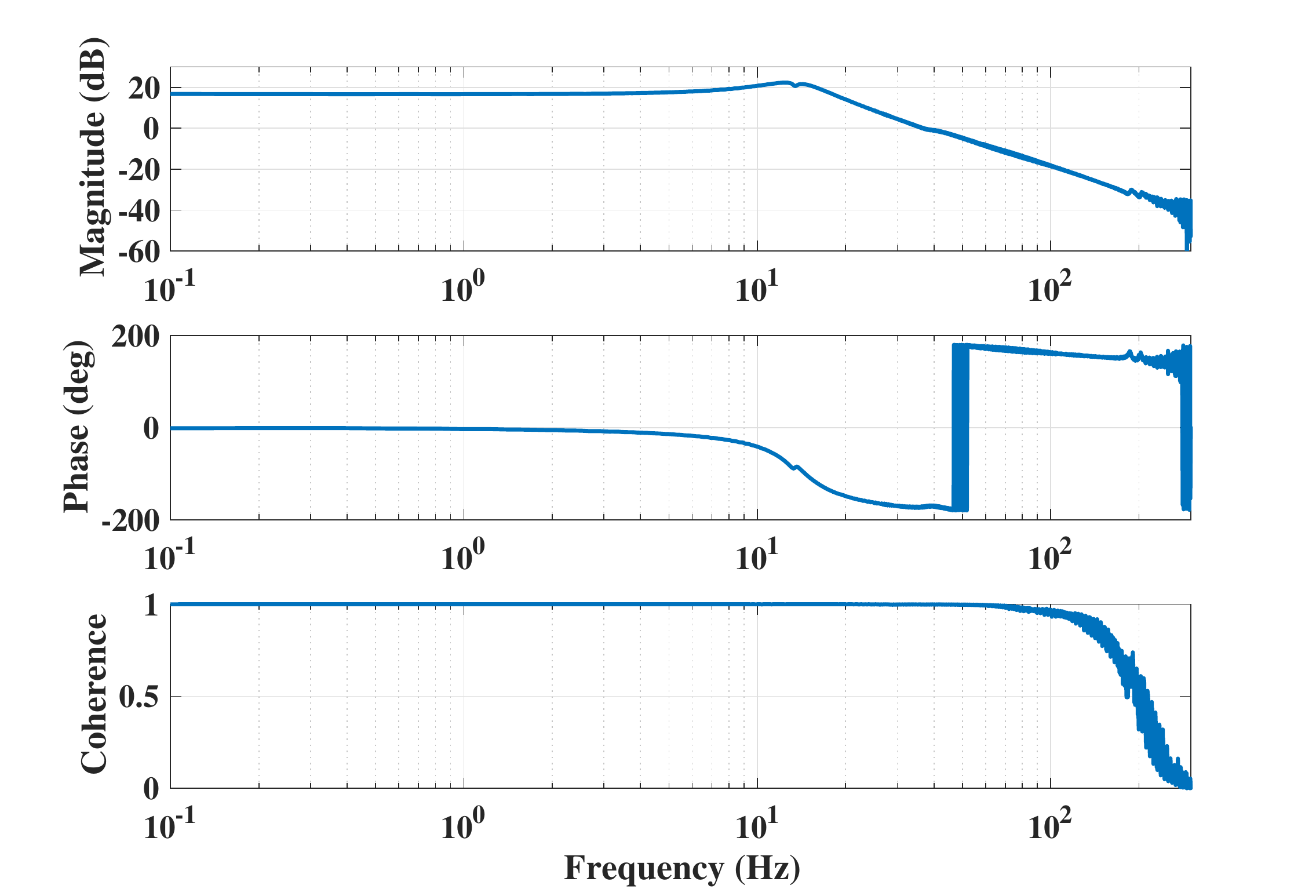}
	\caption{Frequency response data of system as seen from actuator '1A' to position of mass '3' attached to same actuator.}
	\label{fig:frf}	
\end{figure}

The precision planar positioning system shown in Fig. \ref{fig:setup} is used for implementation of all 8 controllers and performance analysis. For sake of simplicity, only one of the actuators (1A) is considered and the position of mass `3' attached to same actuator controlled resulting in a SISO system. All the controllers are implemented using FPGA on NI CompactRIO system to achieve fast real-time control with a sampling frequency of 10 KHz. LM388 linear power amplifier is used for powering the actuator, while Mercury M2000 linear encoder provides position feedback with resolution of $100\ nm$. The frequency response data of the system is obtained by applying a chirp signal and this is shown in Fig. \ref{fig:frf}.

\subsection{Controller design and harmonic analysis}
In the precision industry, bandwidth of motion systems is defined as the cross-over frequency of the open-loop. For the system under consideration, we choose a bandwidth of 150 Hz (942 rad/s). PID is designed for this bandwidth to have a phase margin of $30^\circ$ as below.
\begin{align}
PID=k_p.\left(1+\frac{\omega_i}{s}\right).\left(\frac{s/\omega_d+1}{s/\omega_t+1}\right).\left(\frac{1}{s/\omega_{f}+1}\right)
\end{align}
where $\omega_i = 94.28 rad/s$, $\omega_d = 188.57 rad/s$, $\omega_t = 4714 rad/s$, $\omega_f = 9428 rad/s$ and $k_p = 3.49$.

In the case of PI\textsuperscript{2}D, an additional integrator is to be used to increase the gain of the open-loop at low frequencies to enable better vibration rejection. Since we are interested in the frequency range with a maximum value of 30 Hz, PI\textsuperscript{2}D is designed as below. This immediately results in a loss of nearly $11^\circ$ in phase margin.
\begin{align}
\label{PI2D}
PI^2D = (PID).\left(1+\frac{2\pi 30}{s}\right)
\end{align}

This additional integrator is reset in the case of PI(CI)D and is also used in the band-pass reset design along with the LPF and HPF. In the case of band-pass reset, since we are interested in analysing the effect of the higher order harmonics, the cut-off frequencies of the LPF and HPF are chosen to be 0.01 Hz. With this choice, using IbLPF will result in the attenuation of resetting action and the corresponding harmonics after 0.01 Hz and more importantly in the frequency range of interest (0.5 to 30 Hz), while the opposite is true for IbHPF. In the case of CgLp-PI\textsuperscript{2}D, the CgLp element is added in series before the PI\textsuperscript{2}D which is designed as given in Eqn. (\ref{PI2D}). Although CgLp can provide large phase compensation, it is designed to compensate for the $11^\circ$ phase loss seen due to the introduction of second integrator. The value of $\omega_{r}$ is hence $1317 rad/s$ to ensure $11^\circ$ phase lead from CgLp at bandwidth.

\begin{figure}
	\centering
	\includegraphics[width=0.9\linewidth]{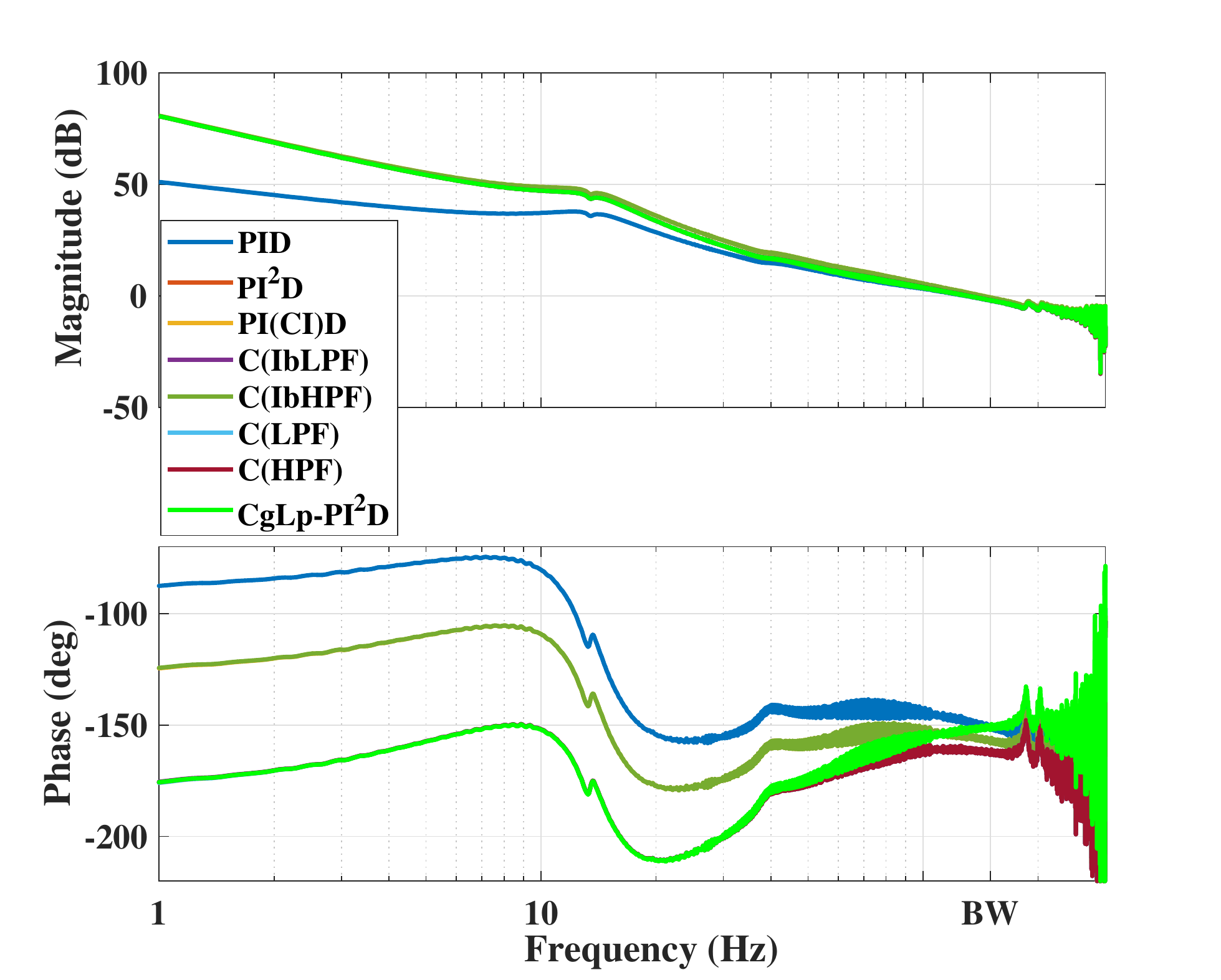}
	\caption{Open-loop of all designed controllers. DF is used for all reset based controllers}
	\label{fig:ol}
\end{figure}

\begin{figure}
	\centering
	\includegraphics[width=0.9\linewidth]{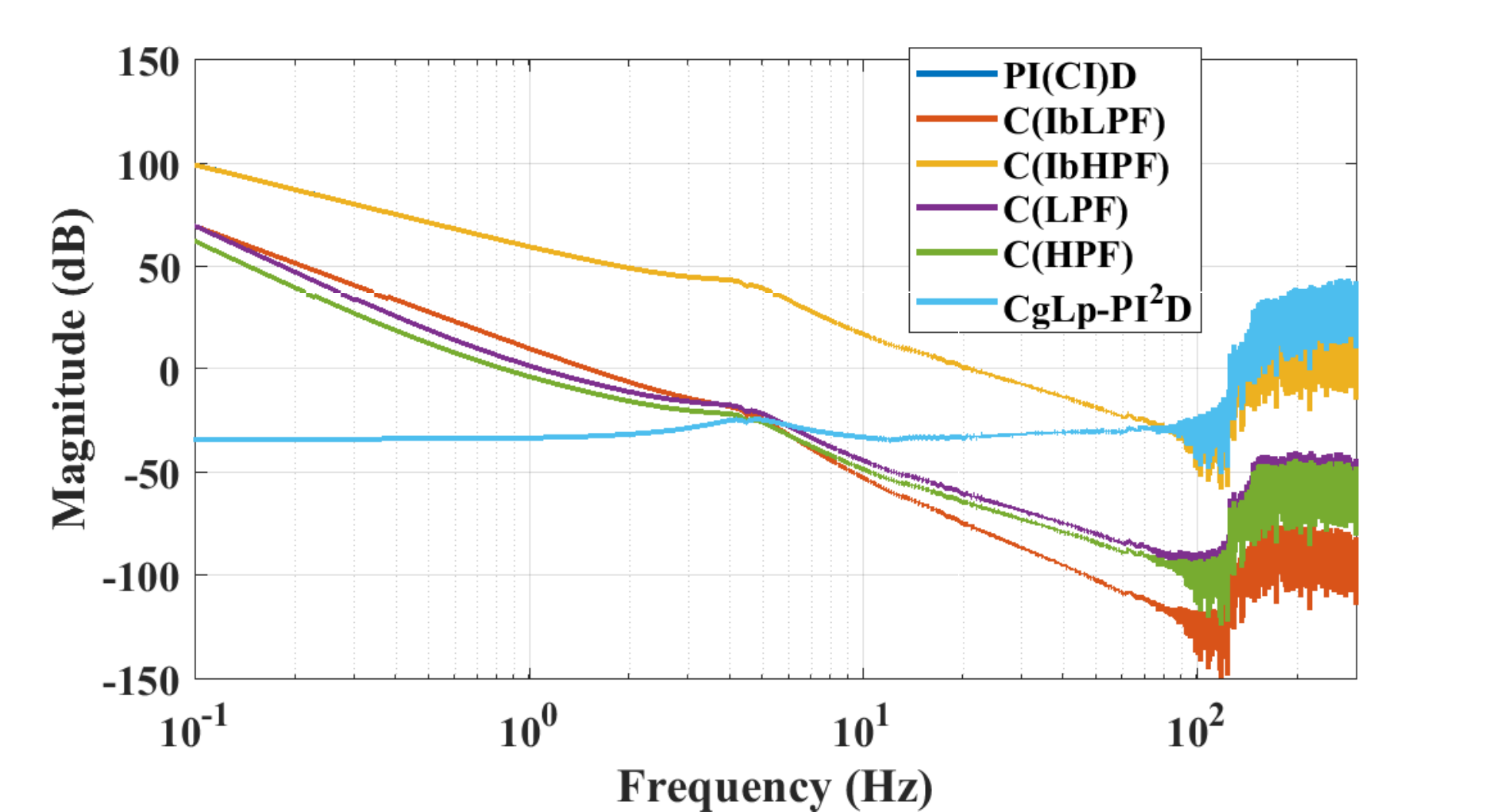}
	\caption{Open-loop third harmonic response for all reset based controllers}
	\label{fig:3ol}
\end{figure}

The open-loop frequency response of the system with all the designed 8 controllers is shown in Fig. \ref{fig:ol}. In the case of controllers using reset, DF (Eqn. \ref{eq:DF}) is used to obtain the behaviour. The 3rd harmonic of the reset based controllers are also obtained using Eqn. (\ref{eq:HOSIDF}) and plotted in Fig. \ref{fig:3ol}. From the two figures, it is clear that the first harmonic frequency gain behaviour of all reset based controllers is the same as that of PI\textsuperscript{2}D. In the case of band-pass reset configurations, phase margin increase is seen in reference to PI\textsuperscript{2}D for the configuration C(IbHPF). However, it is important to note that even in cases where the first harmonic of 2 controllers match in both gain and phase, the third harmonic gain characteristic is different and this provides us with the perfect opportunity to analyse the effect of these harmonics.

\subsection{Results and analysis}
\begin{figure}
	\centering
	\includegraphics[width=0.9\linewidth]{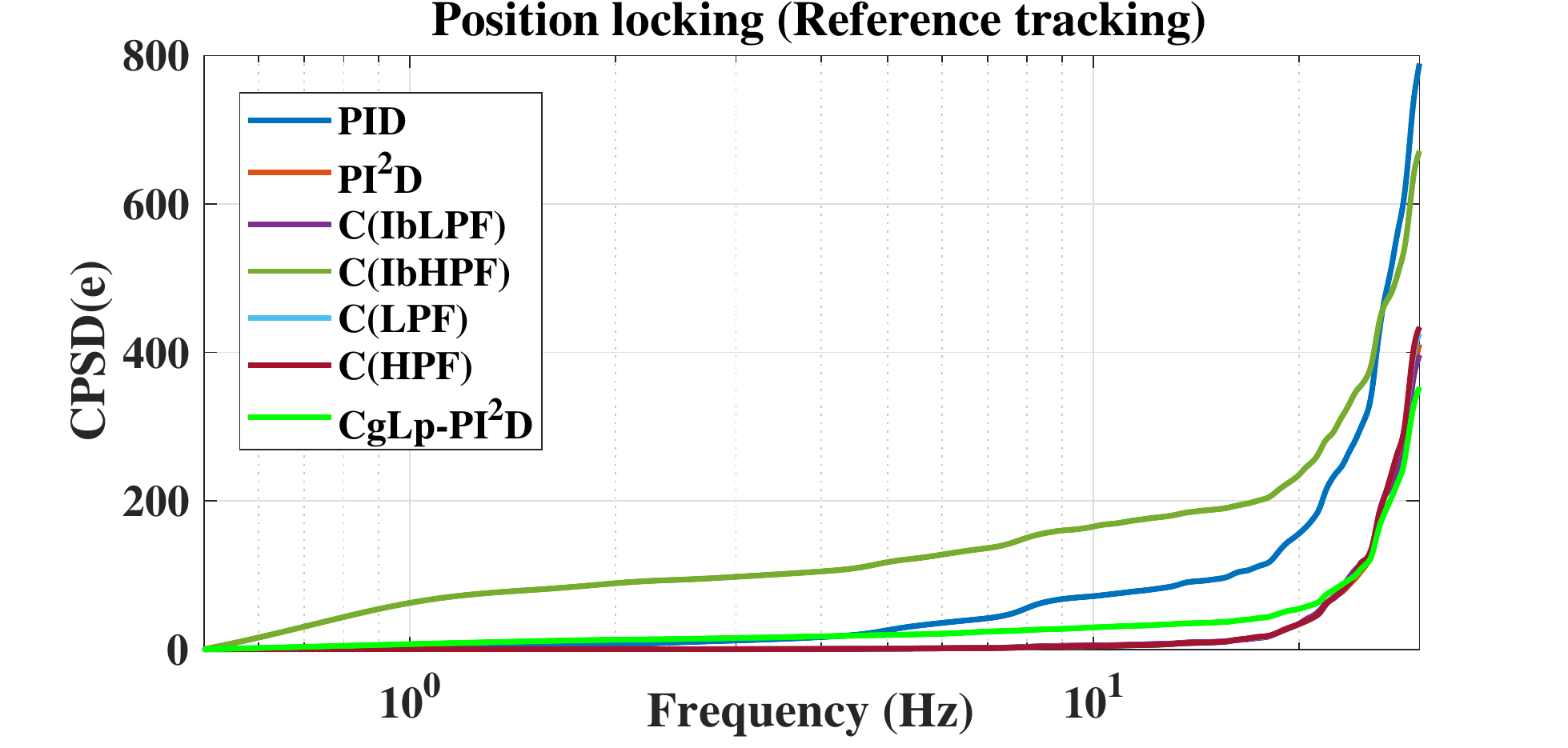}
	\caption{Cumulative PSD for reference tracking}
	\label{fig:rt}
\end{figure}

\begin{figure}
	\centering
	\includegraphics[width=0.9\linewidth]{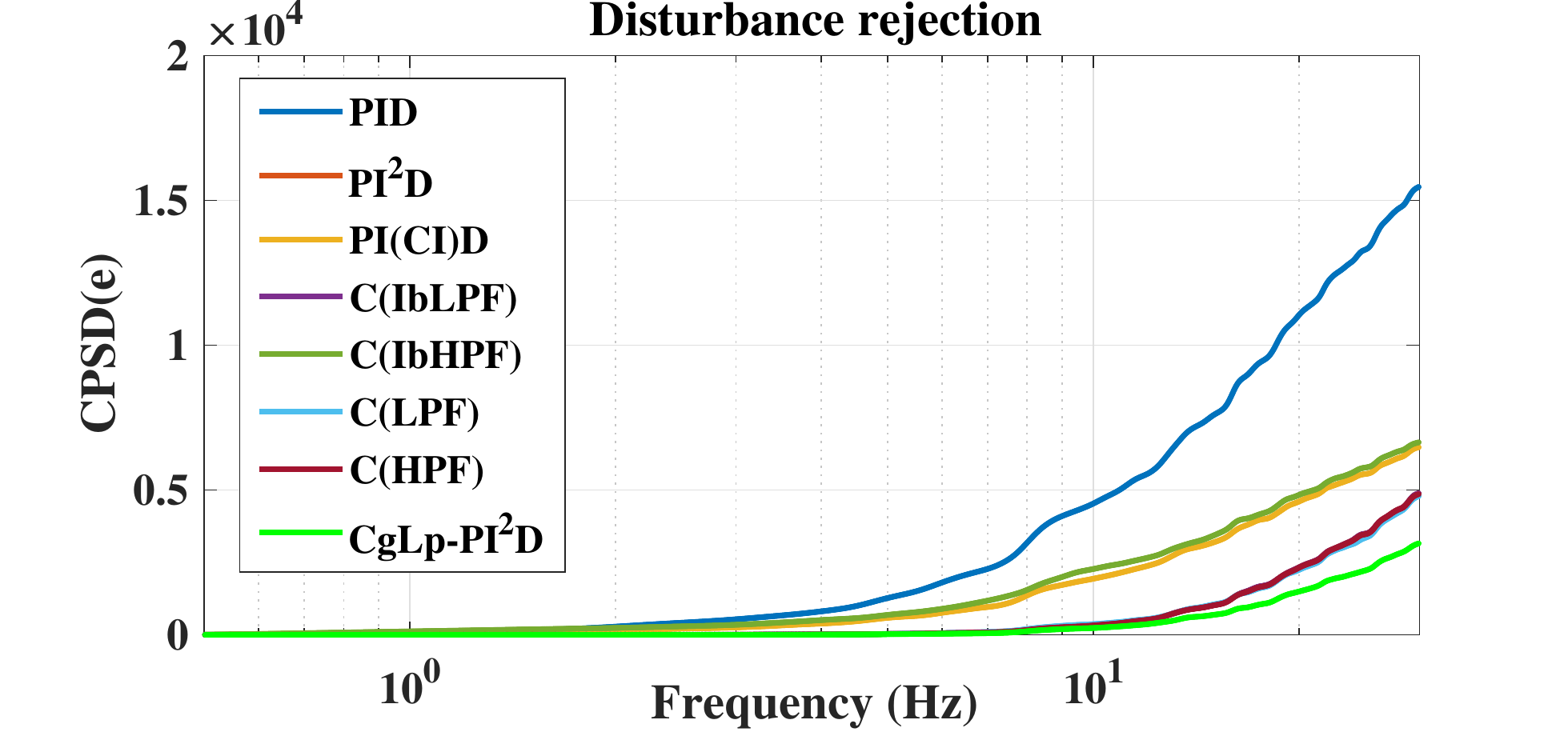}
	\caption{Cumulative PSD for disturbance rejection}
	\label{fig:dr}
\end{figure}

\begin{figure}
	\centering
	\includegraphics[width=0.9\linewidth]{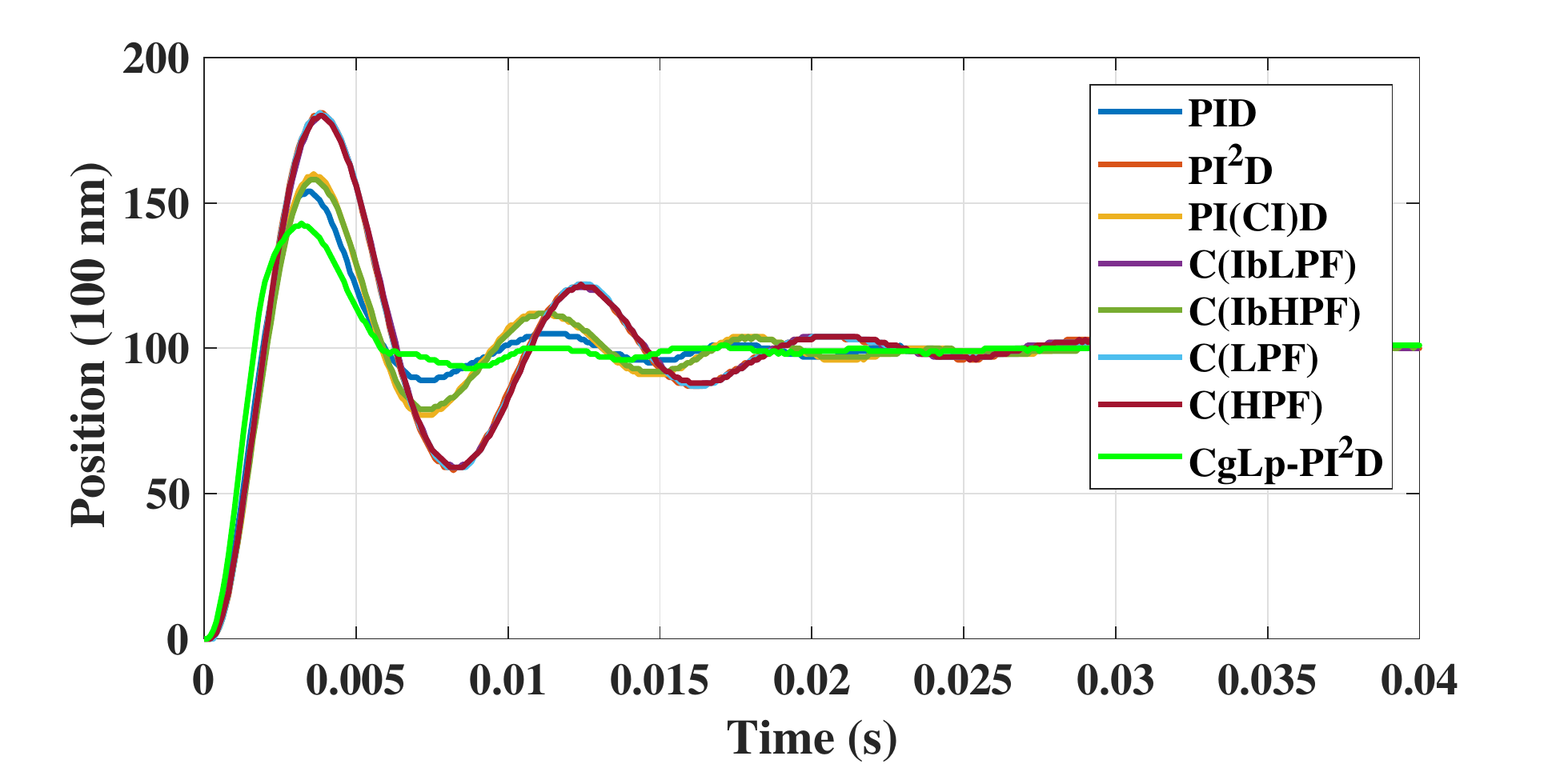}
	\caption{Step responses of all controllers. Step responses of PI\textsuperscript{2}D, C(IbLPF), C(LPF) and C(HPF) overlap. Step responses of PI(CI)D and C(IbHPF) also overlaps.}
	\label{fig:step}
\end{figure}

As noted in Section. \ref{Prelim}, vibration can enter the system either as a disturbance force directly acting on the mass whose position is being controlled or it may be seen as an unknown reference which needs to be followed as in the case of position locking. In the case of linear controller design, the analysis of one of these scenarios would suffice. However, since reset is nonlinear, the performance of all controllers for both these scenarios is tested. For disturbance rejection, a sinusoidal disturbance force is separately added at frequencies of 0.5, 10 and 20 Hz for a period of 60 seconds in all cases and the rms and maximum error noted. The same is also done in the case of reference tracking. Since this is done in the scenario of position locking where reference changes in regard to incoming vibrations, a feedforward controller cannot be used to improve precision tracking performance. White noise passed through a band-pass filter in range 0.5 to 30 Hz is also input as both disturbance and reference separately and CPSD of the error obtained for performance analysis. The rms and maximum errors are provided in Table. \ref{tab::measurementresult}, while the cpsd of errors is shown in Figs. \ref{fig:rt} and \ref{fig:dr}. The step responses of all the controllers are also shown in Fig. \ref{fig:step}.

\begin{table*}
	\centering
	\begin{tabular}{|c||c|c||c|c||c|c||c|c||c|c||c|c|}
		\hline   
		& \multicolumn{6}{c||}{Reference tracking} & \multicolumn{6}{c|}{Disturbance rejection} \\ \hline
		& \multicolumn{2}{c||}{\textbf{0.5 Hz}} & \multicolumn{2}{c||}{\textbf{10 Hz}} & \multicolumn{2}{c||}{\textbf{20 Hz}} & \multicolumn{2}{c||}{\textbf{0.5 Hz}} & \multicolumn{2}{c||}{\textbf{10 Hz}} & \multicolumn{2}{c|}{\textbf{20 Hz}}                  \\ \hline 
		& \multicolumn{1}{c|}{\textbf{RMS}} & \textbf{Max} & \multicolumn{1}{c|}{\textbf{RMS}} & \textbf{Max} & \multicolumn{1}{c|}{\textbf{RMS}} & \textbf{Max} & \multicolumn{1}{c|}{\textbf{RMS}} & \textbf{Max} & \multicolumn{1}{c|}{\textbf{RMS}} & \textbf{Max} & \multicolumn{1}{c|}{\textbf{RMS}} & \textbf{Max} \\ \hline
		\textbf{PID} & 5.04 & 15 & 29.95 & 56 & 81.90 & 121 & 20.42 & 31 & 318.82 & 452 & 418.31 & 593 \\
		\textbf{PI\textsuperscript{2}D} & 4.04 & 12 & 12.49 & 26 & 53.71 & 85 & 0.62 & 1 & 123.77 & 176 & 278.06 & 395 \\
		\textbf{PI(CI)D} & 5.11 & 15 & 23.03 & 66 & 54.51 & 111 & 11.49 & 48 & 203.93 & 622 & 279.87 & 510 \\
		\textbf{C(IbLPF)} & 4.03 & 13 & 12.48 & 25 & 53.80 & 85 & 0.62 & 1 & 123.74 & 176 & 278.03 & 395\\
		\textbf{C(IbHPF)} & 5.35 & 16 & 23.08 & 64 & 55.33 & 112 & 11.07 & 51 & 203.14 & 619 & 279.66 & 508 \\
		\textbf{C(LPF)} & 4.04 & 12 & 12.48 & 25 & 53.82 & 85 & 0.62 & 1 & 123.74 & 176 & 277.93 & 395 \\
		\textbf{C(HPF)} & 4.05 & 13 & 12.49 & 25 & 53.88 & 85 & 0.62 & 1 & 123.78 & 176 & 278.12 & 396 \\
		\textbf{CgLp-PI\textsuperscript{2}D} & 4.37 & 12 & 11.30 & 28 & 44.32 & 71 & 0.81 & 2 & 134.49 & 198 & 248.43 & 337\\
		\hline
	\end{tabular}
	\caption{Measurement error results in 100 nm.}
	\label{tab::measurementresult}
\end{table*}

From the results, it is clear that increased open-loop gain with PI\textsuperscript{2}D results in better performance. While more phase margin is achieved with PI(CI)D, the effect of higher order harmonics completely nullifies this increase. In the case of band-pass reset configurations, the results match either that of PI\textsuperscript{2}D or PI(CI)D even though the third harmonic of these configurations are significantly different in a large frequency range of interest. Further, it is also clear that CgLp-PI\textsuperscript{2}D provides the best overall performance in both disturbance rejection and reference tracking from the context of vibration rejection. It is hypothesized that although in the case of band-pass reset, some cases result in a reduction of third harmonic and subsequently other higher harmonics as well, this reduction is not as significant as seen in the case of using CgLp, especially at lower frequency values. However, even in the case after around 6 Hz, where the third harmonic magnitude of CgLp-PI\textsuperscript{2}D is comparable or higher than that of other band-pass reset configurations, it is seen that the performance is better or comparable. This can be explained by the fact that with the use of CgLp, the reduction of phase margin due to the use of second integrator is completely compensated. This is further confirmed from the step responses where CgLp-PI\textsuperscript{2}D has the lowest overshoot and fastest settling time. In the case of band-pass reset configurations, the phase margin loss is only partially recovered through the use of reset. Hence, combined with the reduction of higher order harmonics and phase margin compensation, CgLp-PI\textsuperscript{2}D provides the best performance. 

\section{Conclusions}
\label{Conclude}

Floor vibrations are a common problem in the precision industry where high accuracy and bandwidths are required. The limitations of linear control encourage us to look towards industry compatible nonlinear techniques with reset control being one such controller. However, resetting action introduces harmonics into the system and can result in performance deterioration. A pure describing function based approach is insufficient.

In this paper, we look at the analysis of reset based controllers for rejecting vibrations with the use of HOSIDF for visualising the higher order harmonics in the open-loop. It is hypothesized that the performance deviation from a pure DF analysis is caused by these higher order harmonics and subsequent attenuation of the same would result in improved performance. To this end, we propose two different methods to achieve phase margin increase along with a reduction of higher order harmonics. Different controllers resulting in similar open-loop gain of first harmonic, but with significantly different higher harmonic gains are designed and tested on a precision positioning setup.

Results are obtained for both unknown reference tracking and disturbance rejection. And it is seen that CgLp-PI\textsuperscript{2}D which overall has the least magnitude of higher order harmonics in region of interest performs best. Although this provides some evidence to the hypothesis that attenuation of harmonics results in better performance, rigorous mathematical relations between open-loop and closed-loop frequency response is required for these systems for optimal design of these controllers. Another important observation is from Fig. \ref{fig:dr}, where CgLp-PI\textsuperscript{2}D performs significantly better than all controllers in all frequency ranges as opposed to its performance in Fig. \ref{fig:rt}, which again shows that the relation between open-loop and all different closed-loop responses is required, so that the full potential of reset systems can be tapped.

\bibliography{thebibliography}

\end{document}